\documentclass{sig-alternate}

\usepackage{url}
\usepackage{graphicx}
\usepackage{color}

\definecolor{darkgreen}{rgb}{0.1,0.6,0.1}

\begin{document}

\title{Recipe recommendation using ingredient networks}

\numberofauthors{3}  

\author{
\alignauthor
Chun-Yuen Teng\\
       \affaddr{School of Information}\\
       \affaddr{University of Michigan}\\
       \affaddr{Ann Arbor, MI, USA}\\
       \email{chunyuen@umich.edu}
\alignauthor
Yu-Ru Lin\\
       \affaddr{IQSS, Harvard University}\\
       \affaddr{CCS, Northeastern University}\\
       \affaddr{Boston, MA}\\
       \email{yuruliny@gmail.com}
\alignauthor Lada A. Adamic\\
       \affaddr{School of Information}\\
       \affaddr{University of Michigan}\\
       \affaddr{Ann Arbor, MI, USA}\\
       \email{ladamic@umich.edu}
}

\date{30 July 1999}

\maketitle
\begin{abstract}
The recording and sharing of cooking recipes, a human activity dating back thousands of years, naturally became an early and prominent social use of the web.  
The resulting online recipe collections are repositories of ingredient combinations and cooking methods whose large-scale and variety yield interesting insights about both the fundamentals of cooking and user preferences. 
At the level of an individual ingredient we measure whether it tends to be essential or can be dropped or added, and whether its quantity can be modified. We also construct two types of networks to capture the relationships between ingredients. The complement network captures which ingredients tend to co-occur frequently, and is composed of two large communities: one savory, the other sweet. The substitute network, derived from user-generated suggestions for modifications, can be decomposed into many communities of functionally equivalent ingredients, and captures users' preference for healthier variants of a recipe. Our experiments reveal that recipe ratings can be well predicted with features derived from combinations of ingredient networks and nutrition information.
\end{abstract}




 \section{Introduction}
The web enables individuals to collaboratively share knowledge and recipe websites are one of the earliest examples of collaborative knowledge sharing on the web.
Allrecipes.com, the subject of our present study, was founded in 1997, years ahead of other collaborative websites such as the Wikipedia. 
Recipe sites thrive because individuals are eager to share their recipes, from family recipes that had been passed down for generations, to new concoctions that they created that afternoon, having been motivated in part by the ability to share the result online. Once shared, the recipes are implemented and evaluated by other users, who supply ratings and comments. 

The desire to look up recipes online may at first appear odd given that tombs of printed recipes can be found in almost every kitchen. The Joy of Cooking~\cite{rombauer1997joy} alone contains 4,500 recipes spread over 1,000 pages. There is, however, substantial additional value in online recipes, beyond their accessibility. While the Joy of Cooking contains a single recipe for Swedish meatballs, Allrecipes.com hosts ``Swedish Meatballs I'', ``II'', and ``III'', submitted by different users, along with 4 other variants, including ``The Amazing Swedish Meatball''. Each variant has been reviewed, from 329 reviews for ``Swedish Meatballs I" to 5 reviews for ``Swedish Meatballs III". The reviews not only provide a crowd-sourced ranking of the different recipes, but also many suggestions on how to modify them, e.g. using ground turkey instead of beef, skipping the ``cream of wheat'' because it is rarely on hand, etc.  

The wealth of information captured by online collaborative recipe sharing sites is revealing not only of the fundamentals of cooking, but also of user preferences.  The co-occurrence of ingredients in tens of thousands of recipes provides information about which ingredients go well together, and when a pairing is unusual. Users' reviews provide clues as to the flexibility of a recipe, and the ingredients within it. Can the amount of cinnamon be doubled? Can the nutmeg be omitted? If one is lacking a certain ingredient, can a substitute be found among supplies at hand without a trip to the grocery store? Unlike cookbooks, which will contain vetted but perhaps not the best variants for some individuals' tastes, ratings assigned to user-submitted recipes allow for the evaluation of what works and what does not. 

In this paper, we seek to distill the collective knowledge and preference about cooking through mining a popular recipe-sharing website. To extract such information, we first parse the unstructured text of the recipes and the accompanying user reviews. We construct two types of networks that reflect different relationships between ingredients, in order to capture users' knowledge about how to combine ingredients. The complement network captures which ingredients tend to co-occur frequently, and is composed of two large communities: one savory, the other sweet. The substitute network, derived from user-generated suggestions for modifications, can be decomposed into many communities of functionally equivalent ingredients, and captures users' preference for healthier variants of a recipe. Our experiments reveal that recipe ratings can be well predicted by features derived from combinations of ingredient networks and nutrition information (with accuracy .792), while most of the prediction power comes from the ingredient networks (84\%).

The rest of the paper is organized as follows. Section~\ref{sec:relwork} reviews the related work. Section~\ref{sec:data} describes the dataset. Section~\ref{sec:complementnet} discusses the extraction of  the ingredient and complement networks and their characteristics. Section~\ref{sec:modification} presents the extraction of recipe modification information, as well as the construction and characteristics of the ingredient substitute network.
Section~\ref{sec:predict} presents our experiments on recipe recommendation and Section~\ref{sec:conclusion} concludes.

\section{Related work}\label{sec:relwork}

Recipe recommendation has been the subject of much prior work. Typically the goal has been to suggest recipes to users based on their past recipe ratings~\cite{svensson2005designing}\cite{forbes2011content} or browsing/cooking history~\cite{uedauser}. The algorithms then find similar recipes based on overlapping ingredients, either treating each ingredient equally~\cite{freyne2010intelligent}  or  by identifying key ingredients~\cite{zhang2008back}.  Instead of modeling recipes using ingredients, Wang et al.~\cite{wang2008substructure} represent the recipes as graphs which are built on ingredients and cooking directions, and they demonstrate that graph representations can be used to easily aggregate Chinese dishes by the flow of cooking steps and the sequence of added ingredients. However, their approach only models the occurrence of ingredients or cooking methods, and doesn't take into account the relationships between ingredients. In contrast, in this paper we incorporate the likelihood of ingredients to co-occur, as well as the potential of one ingredient to act as a substitute for another. 

Another branch of research has focused on recommending recipes based on desired nutritional intake or promoting healthy food choices. Geleijnse et al.~\cite{geleijnse2011personalized} designed a prototype of a personalized recipe advice system, which suggests recipes to users based on their past food selections and nutrition intake. In addition to nutrition information, Kamieth et al.~\cite{kamieth2011adaptive} built a personalized recipe recommendation system based on availability of ingredients and personal nutritional needs. Shidochi et al.~\cite{shidochi2009finding} proposed an algorithm to extract replaceable ingredients from recipes in order to satisfy users' various demands, such as calorie constraints and food availability. Their method identifies substitutable ingredients by matching the cooking actions that correspond to ingredient names. 
However, their assumption that substitutable ingredients are subject to the same processing methods is less direct and specific than extracting substitutions directly from user-contributed suggestions. 

Ahn et al. \cite{ahn2011flavor} and Kinouchi et al \cite{kinouchi2008non} examined networks involving ingredients derived from recipes, with the former modeling ingredients by their flavor bonds, and the latter examining the relationship between ingredients and recipes.  In contrast, we derive direct ingredient-ingredient networks of both compliments and substitutes. We also step beyond characterizing these networks to demonstrating that they can be used to predict which recipes will be successful.   

\vspace{20pt}
\section{Dataset}\label{sec:data}
Allrecipes.com is one of the most popular recipe-sharing websites, where novice and expert cooks alike can upload and rate cooking recipes. It hosts 16 customized international sites
for users to share their recipes in their native languages, of which we study only the main, English, version. Recipes uploaded to the site contain specific instructions on how to prepare a dish: the list of ingredients, preparation steps, preparation and cook time, the number of servings produced, nutrition information, serving directions, and photos of the prepared dish. The uploaded recipes are enriched with user ratings and reviews, which comment on the quality of the recipe, and suggest changes and improvements. In addition to rating and commenting on recipes, users are able to save them as favorites or recommend them to others through a forum. 

We downloaded 46,337 recipes including all information listed from allrecipes.com, including several classifications, such as a region (e.g. the midwest region of US or Europe), the course or meal the dish is appropriate for (e.g.: appetizers or breakfast), and any holidays the dish may be associated with. In order to understand users' recipe preferences, we crawled 1,976,920 reviews which include reviewers' ratings, review text, and the number of users who voted the review as useful.

\subsection{Data preprocessing}
The first step in processing the recipes is identifying the ingredients and cooking methods from the freeform text of the recipe. Usually, although not always, each ingredient is listed on a separate line. To extract the ingredients, we tried two approaches. In the first, we found the maximal match between a pre-curated list of ingredients and the text of the line. However, this missed too many ingredients, while misidentifying others. In the second approach, we used regular expression matching to remove non-ingredient terms from the line and identified the remainder as the ingredient. We removed quantifiers, such as e.g. ``1 lb'' or ``2 cups'', words referring to consistency or temperature, e.g. chopped or cold, along with a few other heuristics, such as removing content in parentheses. For example ``1 (28 ounce) can baked beans (such as Bush's Original\textregistered)" is identified as
``baked beans". By limiting the list of potential terms to remove from an ingredient entry, we erred on the side of not conflating potentially identical or highly similar ingredients, e.g. ``cheddar cheese'', used in 2450 recipes, was considered different from ``sharp cheddar cheese'', occurring in 394 recipes.  

We then generated an ingredient list sorted by frequency of ingredient occurrence and selected the top 1000 common ingredient names as our finalized ingredient list. Each of the top 1000 ingredients occurred in 23 or more recipes, with plain salt making an appearance in 47.3\% of recipes. These ingredients also accounted for 94.9\% of ingredient entries in the recipe dataset. The remaining ingredients were missed either because of high specificity (e.g. yolk-free egg noodle), referencing brand names (e.g. Planters almonds), rarity (e.g. serviceberry), misspellings, or not being a food (e.g. ``nylon netting"). 

The remaining processing task was to identify cooking processes from the directions. We first identified all heating methods using a listing in the Wikipedia entry on cooking ~\cite{wikiCooking}. For example, baking, boiling, and steaming are all ways of heating the food. We then identified mechanical ways of processing the food such as chopping and grinding, and other chemical techniques such as marinating and brining.

\subsection{Regional preferences}
\begin{figure}[t!]
\centering
\includegraphics[trim=2 20 10 10,width=1\columnwidth]{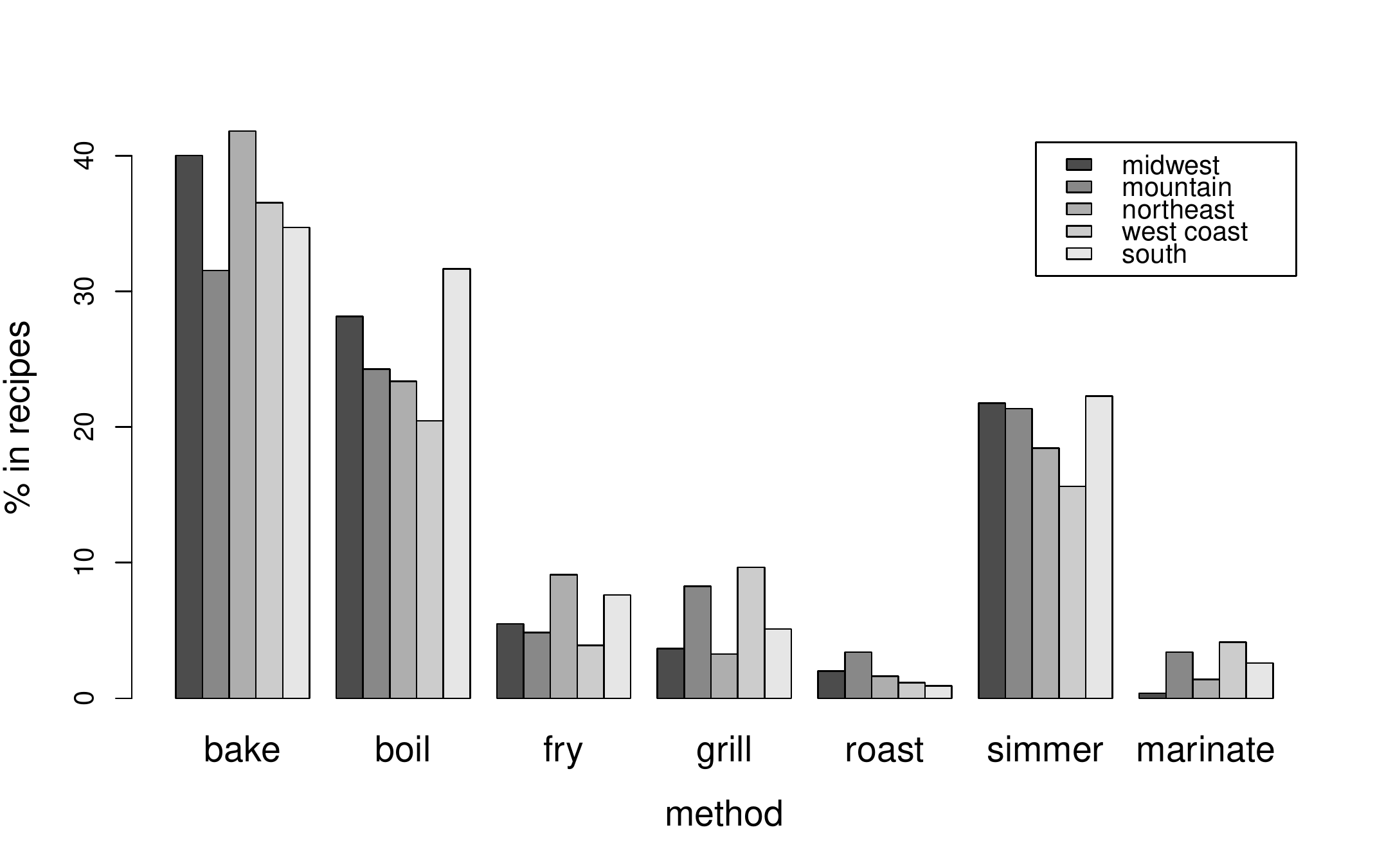}
\caption{ \label{fig:region_method} 
The percentage of recipes by region that apply a specific heating method.}
\vspace{-1em}
\end{figure}

Choosing one cooking method over another appears to be a question of regional taste. 5.8\% of recipes were classified into one of five US regions: Mountain, Midwest, Northeast, South, and West Coast (including Alaska and Hawaii). 
 Figure~\ref{fig:region_method} shows significantly ($\chi^2$ test p-value <~0.001) varying preferences in the different US regions among 6 of the most popular cooking methods. Boiling and simmering, both involving heating food in hot liquids, are more common in the South and Midwest. 
Marinating and grilling are relatively more popular in the West and Mountain regions, but in the West more grilling recipes involve seafood (18/42 = 42\%) relative to other regions combined  (7/106 = 6\%). Frying is popular in the South and Northeast. Baking is a universally popular and versatile technique, which is often used for both sweet and savory dishes, and is slightly more popular in the Northeast and Midwest. Examination of individual recipes reflecting these frequencies shows that these differences in preference can be tied to differences in demographics, immigrant culture and availability of local ingredients, e.g. seafood. 
 
\section{Ingredient complement network}\label{sec:complementnet}
Can we learn how to combine ingredients from the data?
Here we employ the occurrences of ingredients across recipes to distill users' knowledge about combining ingredients. 

We constructed an ingredient complement network based on pointwise mutual information (PMI) defined on pairs of ingredients $(a,b)$:
\[
\mathrm{PMI(a,b)} = log\frac{p(a,b)}{p(a)p(b)},
\]
where
\[
p(a,b)= \frac{\mathrm{\# \:of \:recipes\: containing} \: a \mathrm{\:and} \:b }{\mathrm{\# \:of \:recipes}},
\]
\[
p(a) = \frac{\mathrm{\# \:of \:recipes\: containing} \:a }{\mathrm{\# \:of \:recipes}},
\]
\[
p(b) = \frac{\mathrm{\# \:of \:recipes\: containing} \:b }{\mathrm{\# \:of \:recipes}}.
\]

The PMI gives the probability that two ingredients occur together against the probability that they occur separately. Complementary ingredients tend to occur together far more often than would be expected by chance. 

\begin{figure*}[bth]
\centering
\includegraphics[trim=0 30 0 30,width=1.0\textwidth]{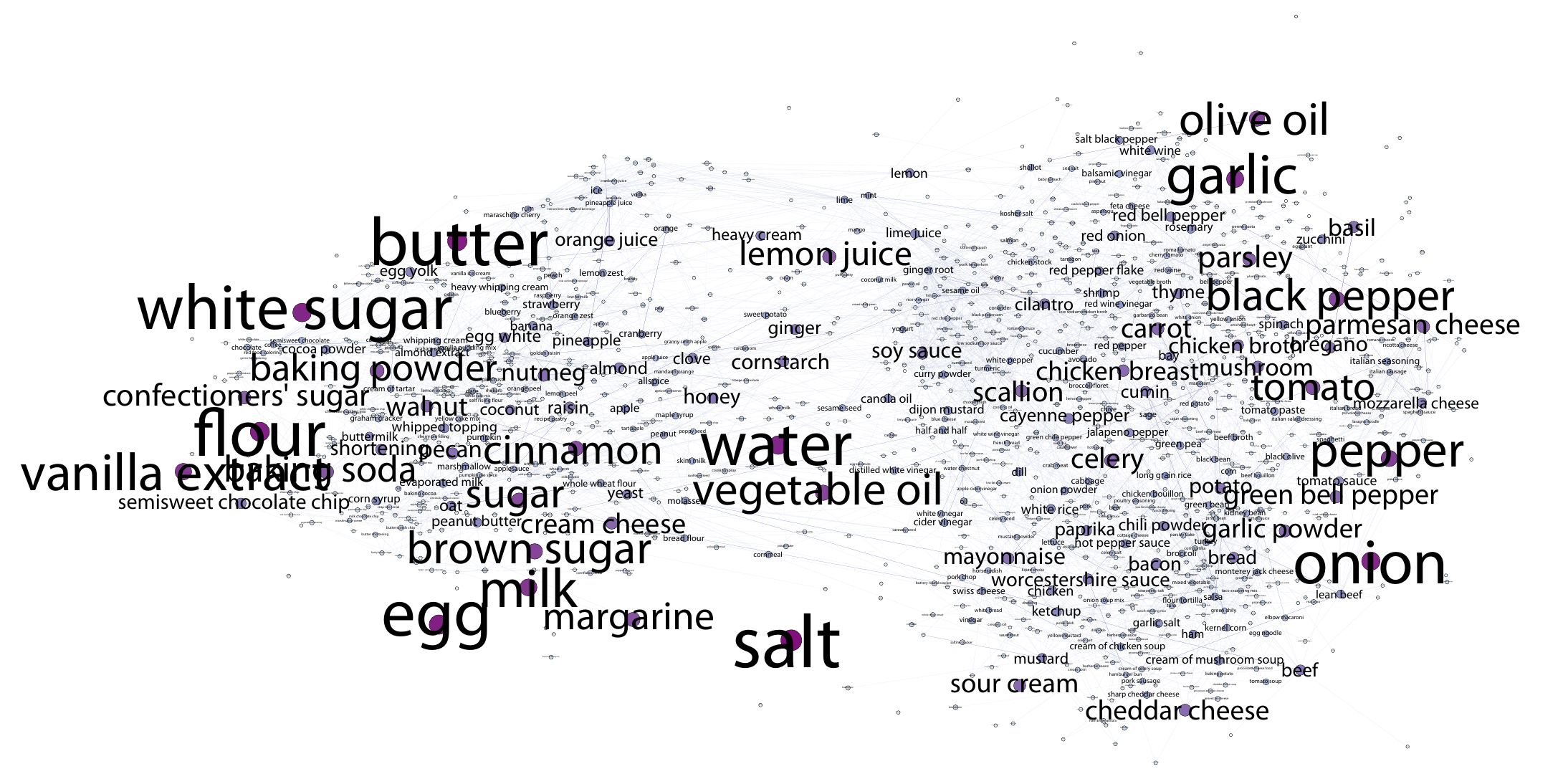}
\caption{Ingredient complement network. Two ingredients share an edge if they occur together more than would be expected by chance and if their pointwise mutual information exceeds a threshold.   \label{fig:complementnet}}
\end{figure*}

Figure~\ref{fig:complementnet} shows a visualization of ingredient complementarity. Two distinct subcommunities of recipes are immediately  apparent: one corresponding to savory dishes, the other to sweet ones. Some central ingredients, e.g. egg and salt, actually are pushed to the periphery of the network. They are so ubiquitous, that although they have many edges, they are all weak, since they don't show particular complementarity with any single group of ingredients. 

We further probed the structure of the complementarity network by applying a network clustering algorithm~\cite{rosvall2008maps}. The algorithm confirmed the existence of two main clusters containing the vast majority of the ingredients. An interesting satellite cluster is that of mixed drink ingredients, which is evident as a constellation of small nodes located near the top of the sweet cluster in Figure~\ref{fig:complementnet}. The cluster includes the following ingredients: lime, rum, ice, orange, pineapple juice, vodka, cranberry juice, lemonade, tequila, etc. 

For each recipe we recorded the minimum, average, and maximum pairwise pointwise mutual information between ingredients. The intuition is that complementary ingredients would yield higher ratings, while ingredients that don't go together would lower the average rating. We found that while the average and minimum pointwise mutual information between ingredients is uncorrelated with ratings, the maximum is very slightly positively correlated with the average rating for the recipe ($\rho = 0.09$, p-value < $10^{-10}$). This suggests that having at least two complementary ingredients very slightly boosts a recipe's prospects, but having clashing or unrelated ingredients does not seem to do harm. 

\section{Recipe modifications}\label{sec:modification}
Co-occurrence of ingredients aggregated over individual recipes reveals the structure of cooking, but tells us little about how flexible the ingredient proportions are, or whether some ingredients could easily be left out or substituted. An experienced cook may know that apple sauce is a low-fat alternative to oil, or may know that nutmeg is often optional, but a novice cook may implement recipes literally, afraid that deviating from the instructions may produce poor results. While a traditional hardcopy cookbook would provide few such hints, they are plentiful in the reviews submitted by users who implemented the recipes, e.g. {\em ``This is a great recipe, but using fresh tomatoes only adds a few minutes to the prep time and makes it taste so much better"}, or another comment about the same salsa recipe {\em ``This is by far the best recipe we have ever come across. We did however change it just a little bit by adding extra onion.''} 

\begin{figure}[t!]
\centering
\includegraphics[trim=0 36 0 40, width=1\columnwidth]{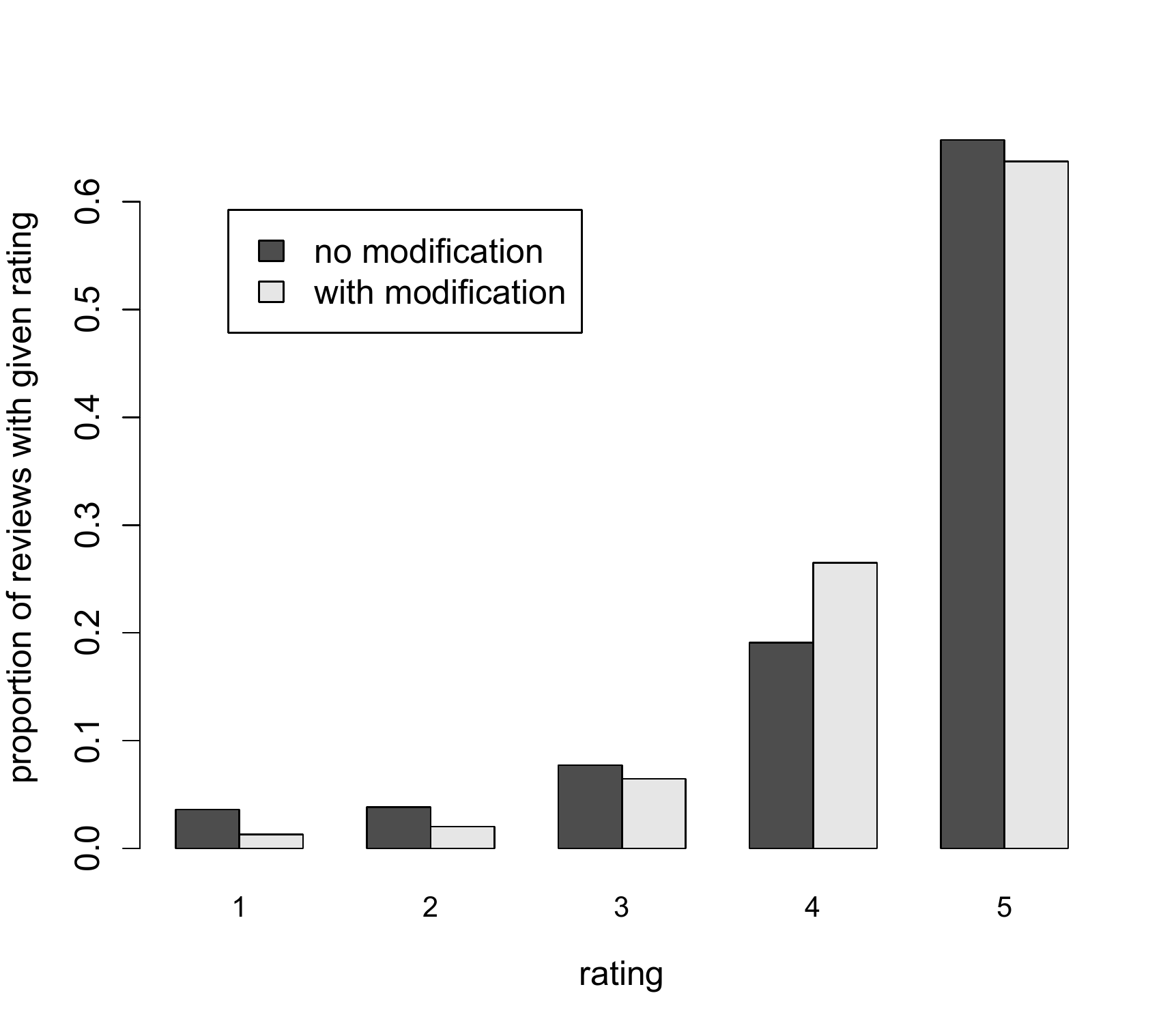}
\caption{The likelihood that a review suggests a modification to the recipe depends on the star rating the review is assigning to the recipe. \label{fig:modornomod}}
\end{figure}

As the examples illustrate, modifications are reported even when the user likes the recipe. In fact, we found that 60.1\% of recipe reviews contain words signaling modification, such as ``add", ``omit", ``instead'', ``extra" and 14 others. Furthermore, it is the reviews that include changes that have a statistically higher average rating (4.49 vs. 4.39, t-test p-value $< 10^{-10}$), and lower rating variance (0.82 vs. 1.05, Bartlett test p-value $< 10^{-10}$), as is evident in the distribution of ratings, shown in Fig.~\ref{fig:modornomod}. This suggests that flexibility in recipes is not necessarily a bad thing, and that reviewers who don't mention modifications are more likely to think of the recipe as perfect, or to dislike it entirely.

In the following, we describe the recipe modifications extracted from user reviews, including adjustment, deletion and addition. We then present how we constructed an ingredient substitute network based on the extracted information.

\subsection{Adjustments}
Some modifications involve increasing or decreasing the amount of an ingredient in the recipe. In this and the following analyses, we split the review on punctuation such as commas and periods.  We used simple heuristics to detect when a review suggested a modification: {\em adding/using more/less} of an ingredient counted as an increase/decrease. Doubling or increasing counted as an increase, while reducing, cutting, or decreasing counted as a decrease. While it is likely that there are other expressions signaling the adjustment of ingredient quantities, using this set of terms allowed us to compare the relative rate of modification, as well as the frequency of increase vs. decrease between ingredients. The ingredients themselves were extracted by performing a maximal character match within a window following an adjustment term. 

\begin{figure}[t!]
\centering
\includegraphics[trim=30 40 0 30,width=1\columnwidth]{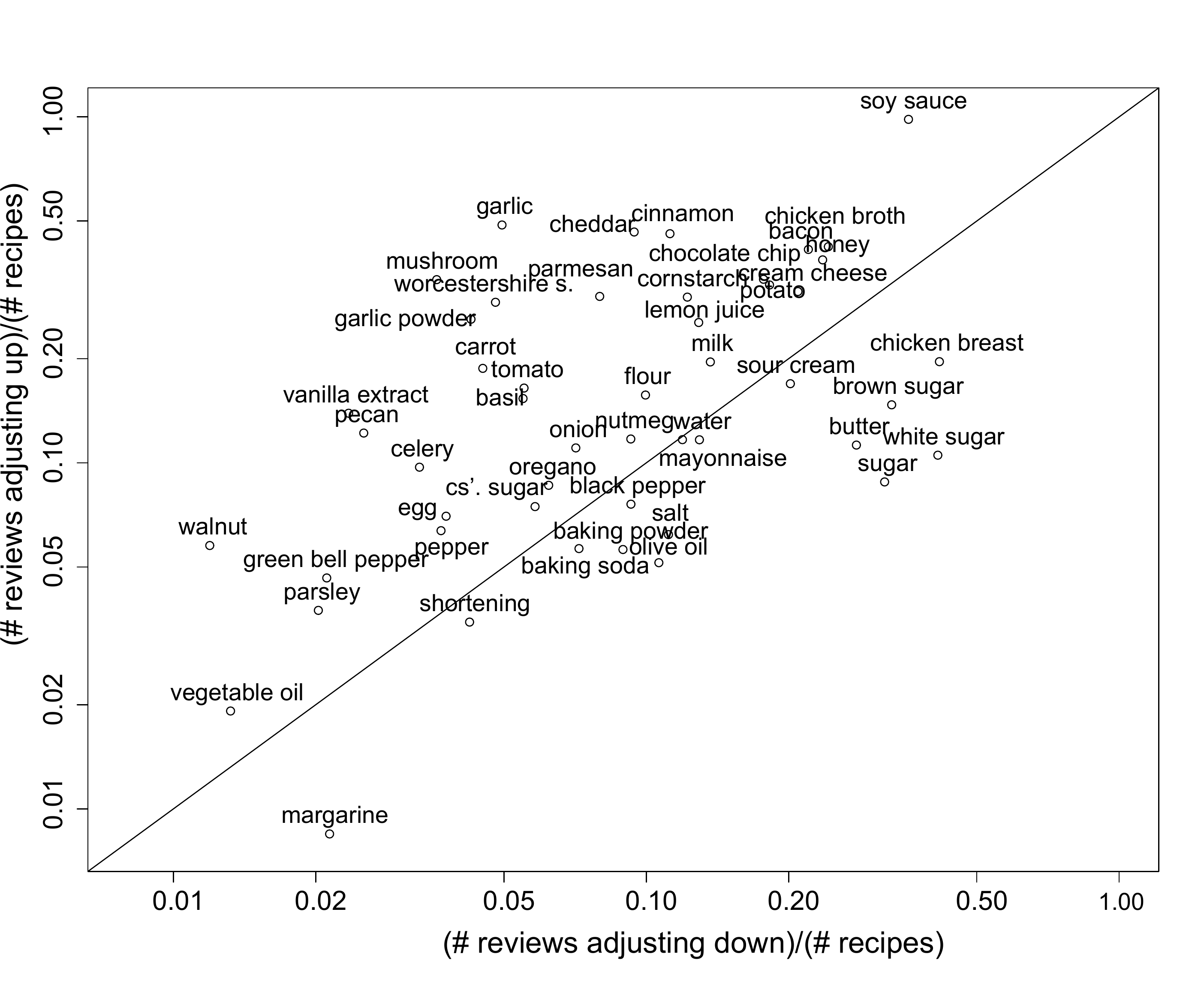}
\caption{Suggested modifications of quantity for the 50 most common ingredients, derived from recipe reviews. The line denotes equal numbers of suggested quantity increases and decreases. \label{fig:updownmod}}
\end{figure}

Figure~\ref{fig:updownmod} shows the ratios of the number of reviews suggesting modifications, either increases or decreases, to the number of recipes that contain the ingredient. Two patterns are immediately apparent. Ingredients that may be perceived as being unhealthy, such as fats and sugars, are, with the exception of vegetable oil and margarine, more likely to be modified, and to be decreased. On the other hand, flavor enhancers such as soy sauce, lemon juice, cinnamon, Worcestershire sauce, and toppings such as cheeses, bacon and mushrooms, are also likely to be modified; however, they tend to be added in greater, rather than lesser quantities. Combined, the patterns suggest that good-tasting but ``unhealthy" ingredients can be reduced, if desired, while spices, extracts, and toppings can be increased to taste.

\subsection{Deletions and additions}
Recipes are also frequently modified such that ingredients are omitted entirely. We looked for words indicating that the reviewer did not have an ingredient (and hence did not use it), e.g. ``had no" and ``didn't have". We further used ``omit/left out/left off/bother with'' as indication that the reviewer had omitted the ingredients, potentially for other reasons. Because reviewers often used simplified terms, e.g. ``vanilla" instead of ``vanilla extract", we compared words in proximity to the action words by constructing 4-character-grams and calculating the cosine similarity between the n-grams in the review and the list of ingredients for the recipe. 

To identify additions, we simply looked for the word ``add", but omitted possible substitutions. For example, we would use ``added cucumber", but not ``added cucumber instead of green pepper", the latter of which we analyze in the following section. We then compared the addition to the list of ingredients in the recipes, and considered the addition valid only if the ingredient does not already belong in the recipe. 

Table~\ref{tab:modcorr} shows the correlation between ingredient modifications. As might be expected, the more frequently an ingredient occurs in a recipe, the more times its quantity has the opportunity to be modified, as is evident in the strong correlation between the the number of recipes the ingredient occurs in and both increases and decreases recommended in reviews. However, the more common an ingredient, the more stable it appears to be. Recipe frequency is negatively correlated with deletions/recipe ($\rho = -0.22$), additions/recipe ($\rho = -0.25$), and  increases/recipe ($\rho = -0.26$).  For example, salt is so essential, appearing in over 21,000 recipes, that we detected only 18 reviews where it was explicitly dropped. In contrast, Worcheshire sauce, appearing in 1,542 recipes, is dropped explicitly in 148 reviews. 

As might also be expected, additions are positively correlated with increases, and deletions with decreases. However, additions and deletions are very weakly negatively correlated, indicating that an ingredient that is added frequently is not necessarily omitted more frequently as well.  

\begin{table}[ht!]
\caption{Correlations between ingredient modifications \label{tab:modcorr}}
\begin{center}
\begin{tabular}{lrrrr}
  \hline
  & addition & deletion & increase & decrease \\ 
  \hline
\# recipes &  0.41 & 0.22 & 0.61 & 0.68 \\ 
  addition &   &-0.15 & 0.79 & 0.11 \\ 
  deletion &    & & 0.09 & 0.58 \\ 
  increase &  & & & 0.39 \\ 
   \hline
\end{tabular}
\end{center}
\vspace{-1em}
\end{table}

\subsection{Ingredient substitute network}\label{sec:substitutenet}
Replacement relationships show whether one ingredient is preferable to another. The preference could be based on taste, availability, or price. Some ingredient substitution tables can be found online\footnote{e.g., http://allrecipes.com/HowTo/common-ingredient-substitutions/detail.aspx}, but are neither extensive nor contain information about relative frequencies of each substitution. Thus, we found an alternative source for extracting replacement relationships -- users' comments, e.g.  {\em ``I replaced the butter in the frosting by sour cream, just to soothe my conscience about all the fatty calories"}. 

To extract such knowledge, we first parsed the reviews as follows: we considered several phrases to signal replacement relationships: ``replace $a$ with $b$'', ``substitute $b$ for $a$'', ``$b$ instead of $a$'', etc, and matched $a$ and $b$ to our list of ingredients. 

We constructed an ingredient substitute network to capture users' knowledge about ingredient replacement. This weighted, directed network consists of ingredients as nodes. We thresholded and eliminated any suggested substitutions that occurred fewer than 5 times. We then determined the weight of each edge by $p(b|a)$, the proportion of substitutions of ingredient $a$ that suggest ingredient $b$. 
For example, 68\% of substitutions for white sugar were to splenda, an artificial sweetener, and hence the assigned weight for the $sugar \rightarrow splenda$ edge is 0.68.  
\begin{figure}[t!]
\centering
\includegraphics[trim=10 30 0 10,width=1.2\columnwidth]{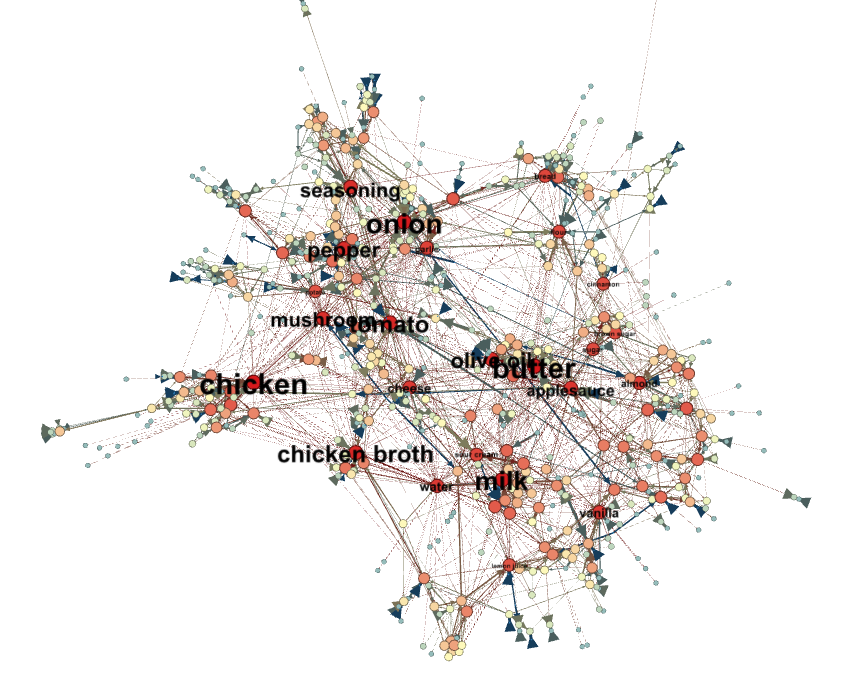}
\caption{Ingredient substitute network. Nodes are sized according to the number of times they have been recommended as a substitute for another ingredient, and colored according to their indegree.  \label{fig:substitutenetwork}}
\vspace{-1em}
\end{figure}

\begin{table}[ht!]
\caption{Clusters of ingredients that can be substituted for one another. A maximum of 5 additional ingredients for each cluster are listed, ordered by PageRank. \label{tab:subclusters}}
\vspace{-1em}
\begin{center}
\begin{tabular}{r|l}
  \hline
 main & other ingredients \\ 
  \hline  \hline
chicken &  turkey, beef, sausage, chicken breast, bacon\\ \hline
olive oil &  butter, apple sauce, oil, banana, margarine \\ \hline
sweet &  yam, potato, pumpkin, butternut squash,  \\
potato & parsnip  \\ \hline
baking  & baking soda, cream of tartar \\ 
powder & \\ \hline
almond & pecan, walnut, cashew, peanut, sunflower s. \\\hline
apple & peach, pineapple, pear, mango, pie filling \\\hline
egg & egg white, egg substitute, egg yolk \\ \hline
tilapia & cod, catfish, flounder, halibut, orange roughy \\ \hline
spinach & mushroom, broccoli, kale, carrot, zucchini \\ \hline
italian & basil, cilantro, oregano, parsley, dill \\ 
seasoning & \\ \hline
cabbage & coleslaw mix, sauerkraut, bok choy\\
 & napa cabbage 
\\ \hline
\end{tabular}
\end{center}
\vspace{-1em}
\end{table}

\begin{figure}[bth]
\centering
\includegraphics[trim=0 20 0 0,width=1\columnwidth]{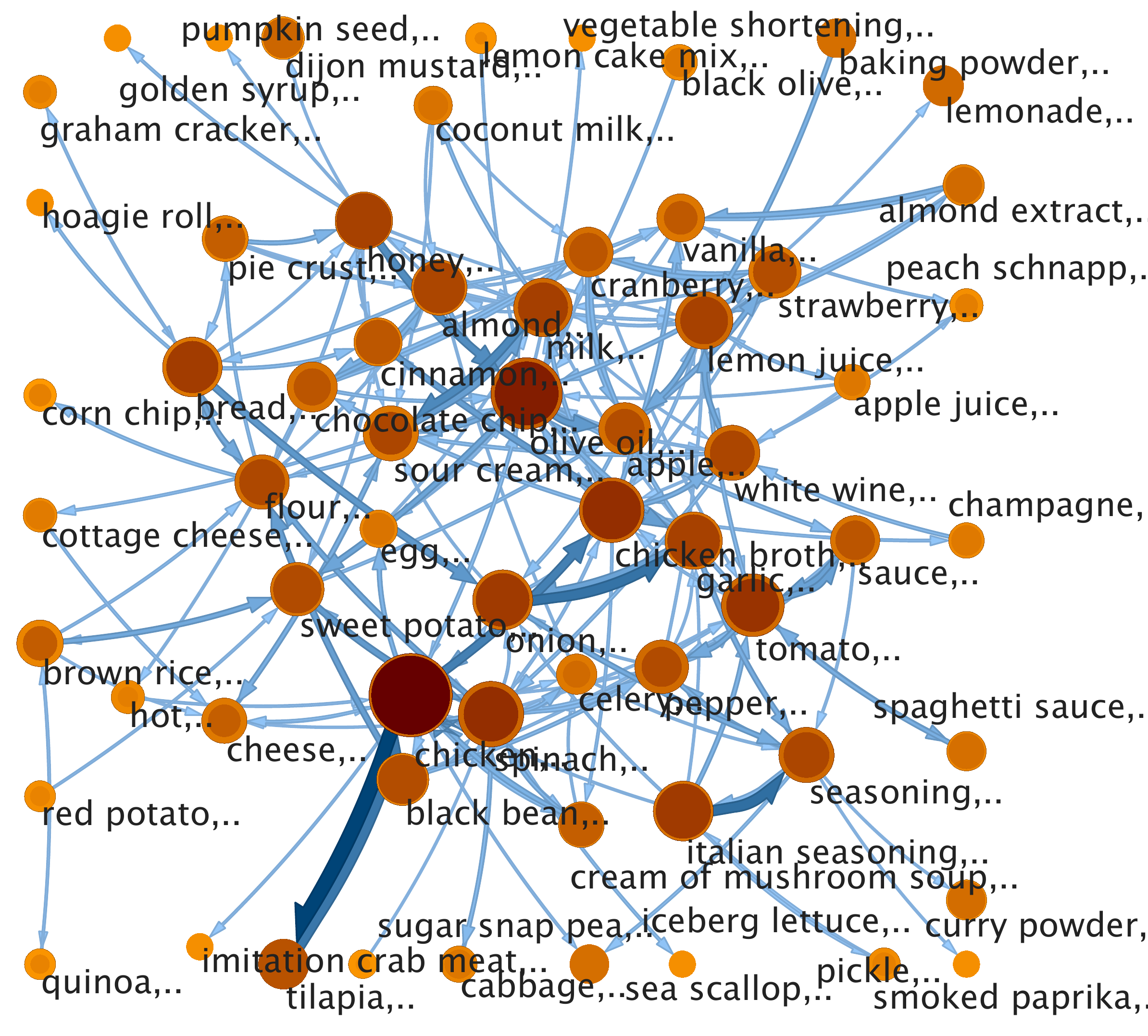}
\caption{\label{fig:subclusters} Ingredient substitution clusters. Nodes represent clusters and edges indicate the presence of recommended substitutions that span clusters. Each cluster represents a set of related ingredients which are frequently substituted for one another.}
\end{figure}

\begin{figure}
\centering
\begin{tabular}{cc}
\includegraphics[width=0.5\columnwidth]{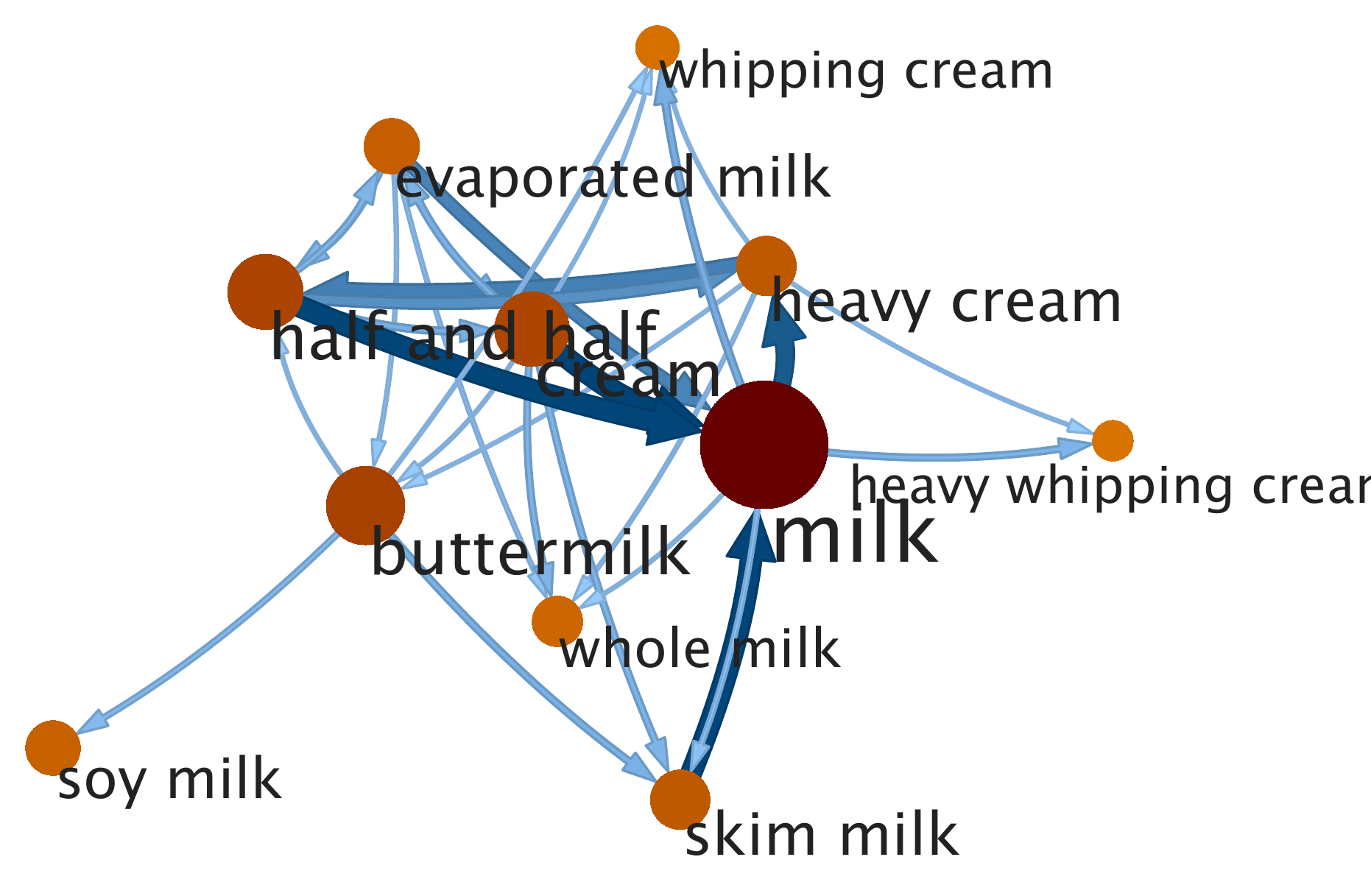} &
\includegraphics[width=0.5\columnwidth]{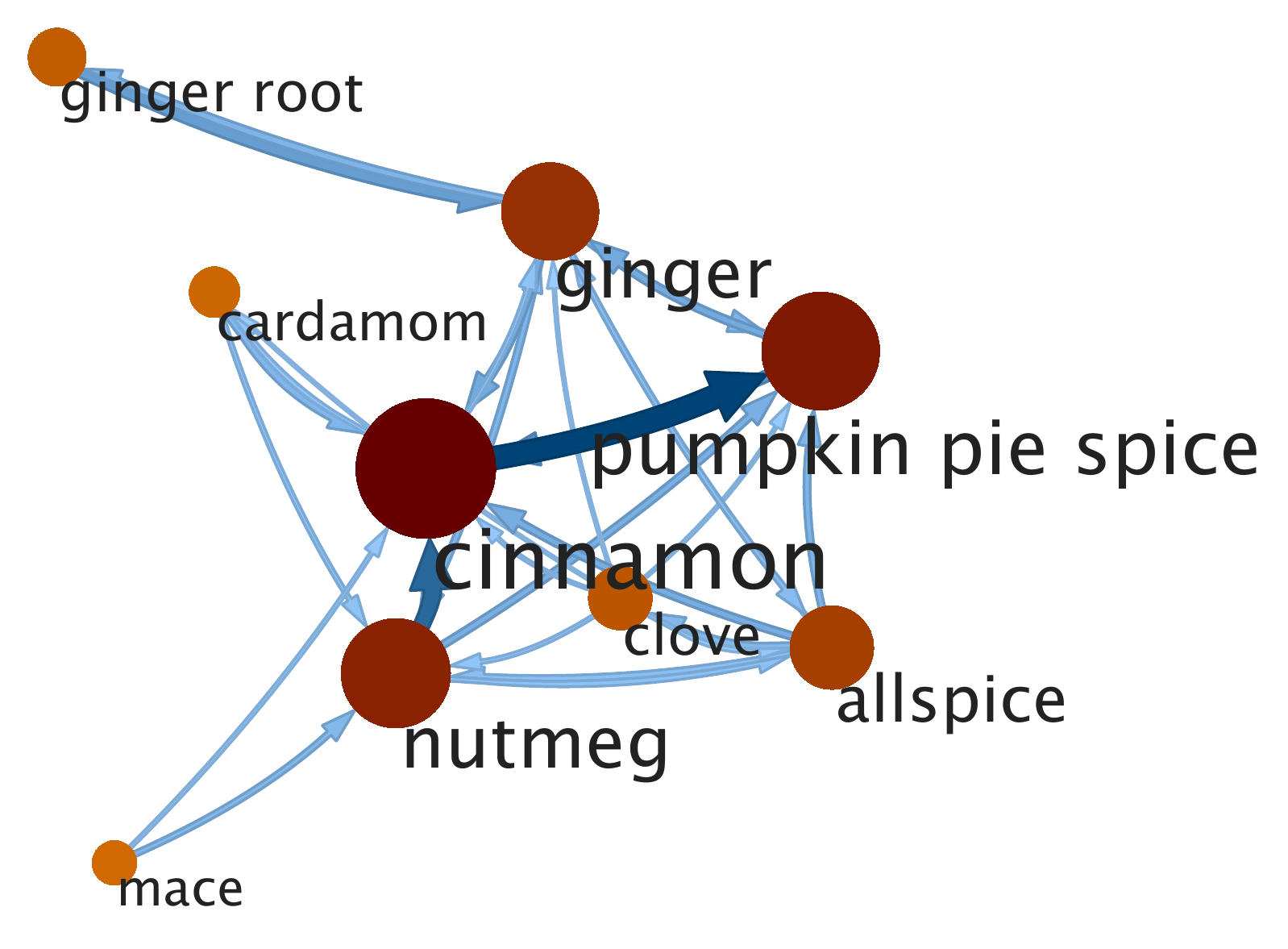}\\
(a) milk substitutes & (b) cinammon substitutes \\
\end{tabular}
\caption{Relationships between ingredients located within two of the clusters from Fig.~\ref{fig:subclusters}.} \label{fig:subcloseup}
\vspace{-1em}
\end{figure}

The resulting substitution network, shown in Figure~\ref{fig:substitutenetwork}, exhibits strong clustering. We examined this structure by applying the map generator tool by Rosvall et al.~\cite{rosvall2008maps}, which uses a random walk approach to identify clusters in weighted, directed networks. The resulting clusters, and their relationships to one another, are shown in Fig.~\ref{fig:subclusters}. The derived clusters could be used when following a relatively new recipe which may not receive many reviews, and therefore many suggestions for ingredient substitutions. If one does not have all ingredients at hand, one could examine the content of one's fridge and pantry and match it with other ingredients found in the same cluster as the ingredient called for by the recipe.  Table~\ref{tab:subclusters} lists the contents of a few such sample ingredient clusters, and Fig.~\ref{fig:subcloseup} shows two example clusters extracted from the substitute network.

\pagebreak
Finally, we examine whether the substitution network encodes preferences for one ingredient over another, as evidenced by the relative ratings of similar recipes, one which contains an original ingredient, and another which implements a substitution. To test this hypothesis, we construct a ``preference network", where one ingredient is preferred to another in terms of received ratings, and is constructed by creating an edge $(a,b)$ between a pair of ingredients, where $a$ and $b$ are listed in two recipes $X$ and $Y$ respectively, if recipe ratings $R_{X}>R_{Y}$. For example, if recipe $X$ includes beef, ketchup and cheese, and recipe $Y$ contains beef and pickles, then this recipe pair contributes to two edges: one from pickles to ketchup, and the other from pickles to cheese. The aggregate edge weights are defined based on PMI. Because PMI is a symmetric quantity ($\mathrm{PMI}(a;b)=\mathrm{PMI}(b;a)$), we introduce a directed PMI measure to cope with the directionality of the preference network:
\[
\mathrm{PMI}(a\to b) = \mathrm{log}\frac{p(a\to b)}{p(a)p(b)},
\]
where
\[
p(a \to b )= \frac{\mathrm{\# \:of \:recipe\: pairs\: from}\: a \mathrm{\:to} \:b }{\mathrm{\# \:of \:recipe \:pairs}},
\]
and $p(a)$, $p(b)$ are defined as in the previous section.

We find high correlation between this preference network and the substitution network ($\rho = 0.72, p < 0.001$). This observation suggests that the substitute network encodes users' ingredient preference, which we use in the recipe prediction task described in the next section.

\section{Recipe recommendation}\label{sec:predict}
We use the above insights to uncover novel recommendation algorithms suitable for recipe recommendations. We use ingredients and the relationships encoded between them in ingredient networks as our main feature sets to predict recipe ratings, and compare them against features encoding nutrition information, as well as other baseline features such as cooking methods, and preparation and cook time. Then we apply a discriminative machine learning method, stochastic gradient boosting trees~\cite{Friedman98additivelogistic}, to predict recipe ratings.

In the experiments, we seek to answer the following three questions.
(1) Can we predict users' preference for a new recipe given the information present in
the recipe?
(2) What are the key aspects that determine users' preference?
(3) Does the structure of ingredient networks help in recipe
recommendation, and how?

\subsection{Recipe Pair Prediction}
The goal of our prediction task is: \textit{given a pair of similar recipes, determine
which one has higher average rating than the other}. This task is
designed particularly to help users with a specific dish or meal in mind, and who
are trying to decide between several recipe options for that dish.

\textbf{Recipe pair data.} The data for this prediction task consists of pairs of
similar recipes. The reason for selecting similar recipes, with high ingredient overlap, is that while apples may be quite comparable to oranges in the context of recipes, especially if one is evaluating salads or desserts, lasagna may not be comparable to a mixed drink. To derive pairs of related recipes, we computed similarity with a cosine similarity between the ingredient 
lists for the two recipes, weighted by the inverse document frequency,
$log(\# \:of \:recipes / \# \:of \:recipes \:containing \:the \:ingredient)$. We
considered only those pairs of recipes whose cosine similarity
exceeded 0.2. 
The weighting is intended to identify higher similarity among recipes sharing more distinguishing ingredients, such as Brussels sprouts, as opposed to recipes sharing very common ones, such as butter. 

A further challenge to obtaining reliable relative rankings of recipes is variance introduced by having different users choose to rate different recipes. In addition, some users might not have a sufficient number of reviews under their belt to have calibrated their own rating scheme. To control for variation introduced by users, we examined recipe pairs where the {\em same} users are rating both recipes and are collectively expressing a preference for one recipe over another.  Specifically, we generated 62,031 recipe pairs ($a,b$) where $rating_i(a)$ > $rating_i(b)$, for at least 10 users $i$, and over 50\% of users who rated both recipe $a$ and recipe $b$. Furthermore, each user $i$ should be an active enough reviewer to have rated at least 8 other recipes.

\textbf{Features.} In the prediction dataset, each observation consists of a set of predictor
variables or features that represent information about two recipes, and the
response variable is a binary indicator of which gets the higher rating on
average. To study the key aspects of recipe information, we constructed
different set of features, including:
\begin{itemize}
\vspace{-.6em}
\item Baseline: This includes cooking methods, such as chopping, marinating, or grilling, and
  cooking effort descriptors, such as preparation time in minutes, as well as the number of
  servings produced, etc. These features are considered as primary information about
  a recipe and will be included in all other
  feature sets described below.
\vspace{-.6em}
\item Full ingredients: We selected up to 1000 popular ingredients to
  build a ``full ingredient list''. In this feature set, each observed
  recipe pair contains a vector with entries indicating whether an ingredient
  from the full list is present in either recipe in the pair.
\vspace{-.6em}
\item Nutrition: This feature set does not include any ingredients but
  only nutrition information such the total caloric content, as well
  as quantities of fats, carbohydrates, etc. 
\vspace{-.6em}
\item Ingredient networks: In this set, we replaced the full ingredient
  list by structural information extracted from different ingredient networks, as
  described in Sections~\ref{sec:complementnet} and~\ref{sec:substitutenet}. Co-occurrence is treated separately as a raw count, and a complementarity, captured by the PMI. 
  \vspace{-.6em}
\item Combined set: Finally, a combined feature set is constructed to
  test the performance of a combination of features, including baseline, nutrition and ingredient networks.
\end{itemize}
\vspace{-.6em}
To build the ingredient network feature set, we extracted the following two types of structural information from
the co-occurrence and substitution networks, as well as the complement
network derived from the co-occurrence information: 

\textit{Network positions} are calculated to represent how a recipe's
ingredients occupy positions within the networks. Such position
measures are likely to inform if a recipe contains any ``popular'' or
``unusual'' ingredients. To calculate the position measures, we first
calculated various network centrality measures, including degree
centrality, betweenness centrality, etc., from the ingredient
networks. A centrality measure can be represented as a vector
$\vec{g}$ where each entry indicates the centrality of an
ingredient. The network position of a recipe, with its full ingredient list represented as a
binary vector $\vec{f}$, can be summarized by
$\vec{g}^{T}\cdot\vec{f}$, i.e., an aggregated centrality measure
based on the centrality of its ingredients.

\textit{Network communities}  provide information about which ingredient is
more likely to co-occur with a group of other ingredients in the
network. A recipe consisting of ingredients that are frequently used
with, complemented by or substituted by certain groups may be predictive of the ratings the recipe will receive. To obtain the network community information,
we applied latent semantic analysis (LSA) on recipes. 
We first factorized each ingredient network, represented by matrix
$W$, 
using singular value decomposition
(SVD). In the matrix $W$, each entry $W_{ij}$ indicates whether ingredient
$i$ co-occurrs, complements or substitues ingredient $j$. 

Suppose
$W_k= U_k\Sigma_kV_k^T$ is a rank-$k$ approximation of $W$, we can
then transform each recipe's full ingredient list using the
low-dimensional representation, $\Sigma_k^{-1} V_k^T \vec{f}$, as
community information within a network. These
low-dimensional vectors, together with the vectors of network
positions, constitute the ingredient network features.

\begin{figure}[t!]\centering
\includegraphics[trim=0 30 0 0,width=1\columnwidth]{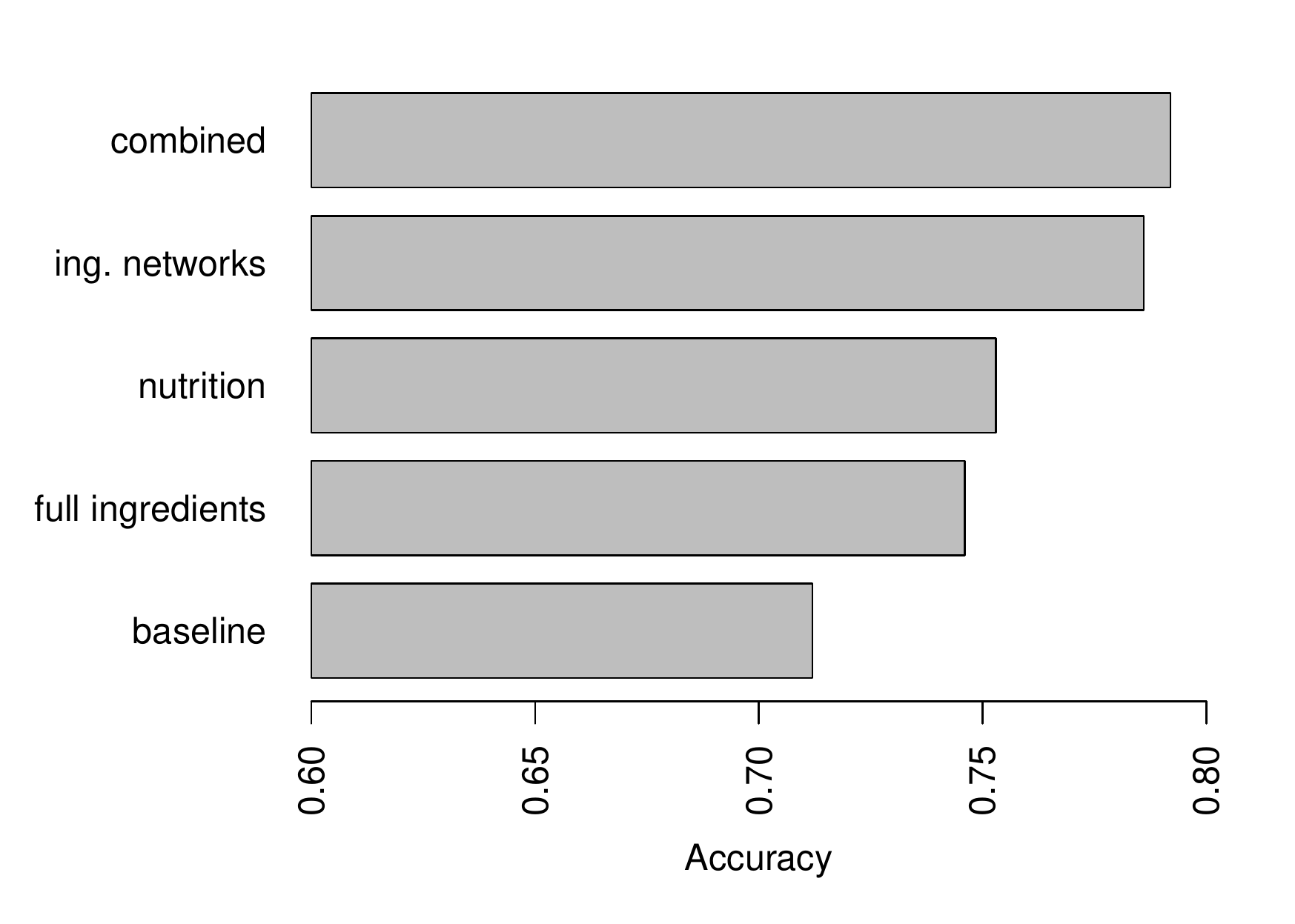}
\caption{\label{fig:predres} Prediction performance. The nutrition
  information and ingredient networks are more effective features than
full ingredients. The ingredient network features lead to impressive performance, close to
the best performance. }  
\vspace{-1em}
\end{figure}

\begin{figure}[t!]\centering
\includegraphics[trim=0 30 0 0,width=1\columnwidth]{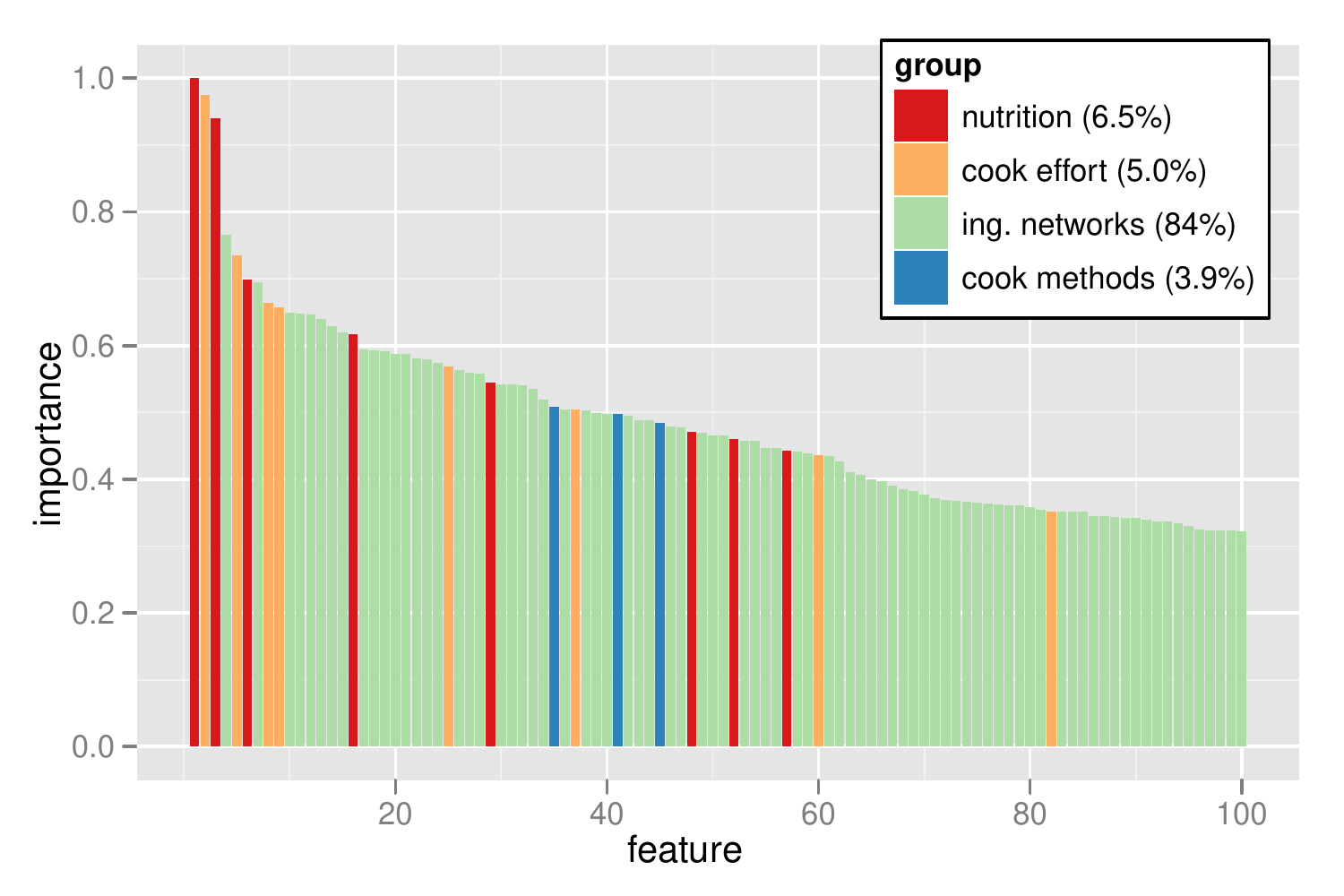}
\caption{\label{fig:relimp} Relative importance of features in the
  combined set. The individual items from nutrition information are very
indicative in differentiating highly rated recipes, while most of the
prediction power comes from ingredient networks.}  
\vspace{-1em}
\end{figure}

\textbf{Learning method.} We applied discriminative machine
learning methods such as support vector machines (SVM)~\cite{cortes1995support} and stochastic
gradient boosting trees~\cite{friedman2002stochastic} to our prediction
problem. Here we report and discuss the detailed results based on the gradient boosting tree model.
Like SVM, the gradient boosting tree model seeks a parameterized
classifier, but unlike SVM that considers all the
features at one time, the boosting tree model considers a set
of features at a time and iteratively combines them according to their
empirical errors. In practice, it not only has competitive performance
comparable to SVM, but can serve as a feature ranking procedure~\cite{lu2006coupling}.

In this work, we fitted a stochastic gradient boosting tree model with 8
terminal nodes under an exponential loss function. The dataset is
roughly balanced in terms of which recipe is the higher-rated one
within a pair. We randomly divided the dataset into a training set
(2/3) and a testing set (1/3). 
The prediction performance is evaluated
based on accuracy, and the feature performance is evaluated in terms
of relative importance~\cite{hastie2005elements}. For each single
decision tree, one of the input variables, $x^j$, is used to partition
the region associated with that node into two subregions in order to
fit to the response values.
The squared relative importance of variable $x^j$ is the sum of such squared improvements over all internal nodes for which it was chosen as the splitting variable, as:
\[
imp(j)=\sum_{k} \hat{i}_k^2 I(\mbox{splits on } x^j)
\]
where $\hat{i}_k^2$ is the empirical improvement by the $k$-th node splitting on
$x^j$ at that point.
\subsection{Results}

\begin{figure}[!]\centering
\includegraphics[trim=0 30 0 0,width=1\columnwidth]{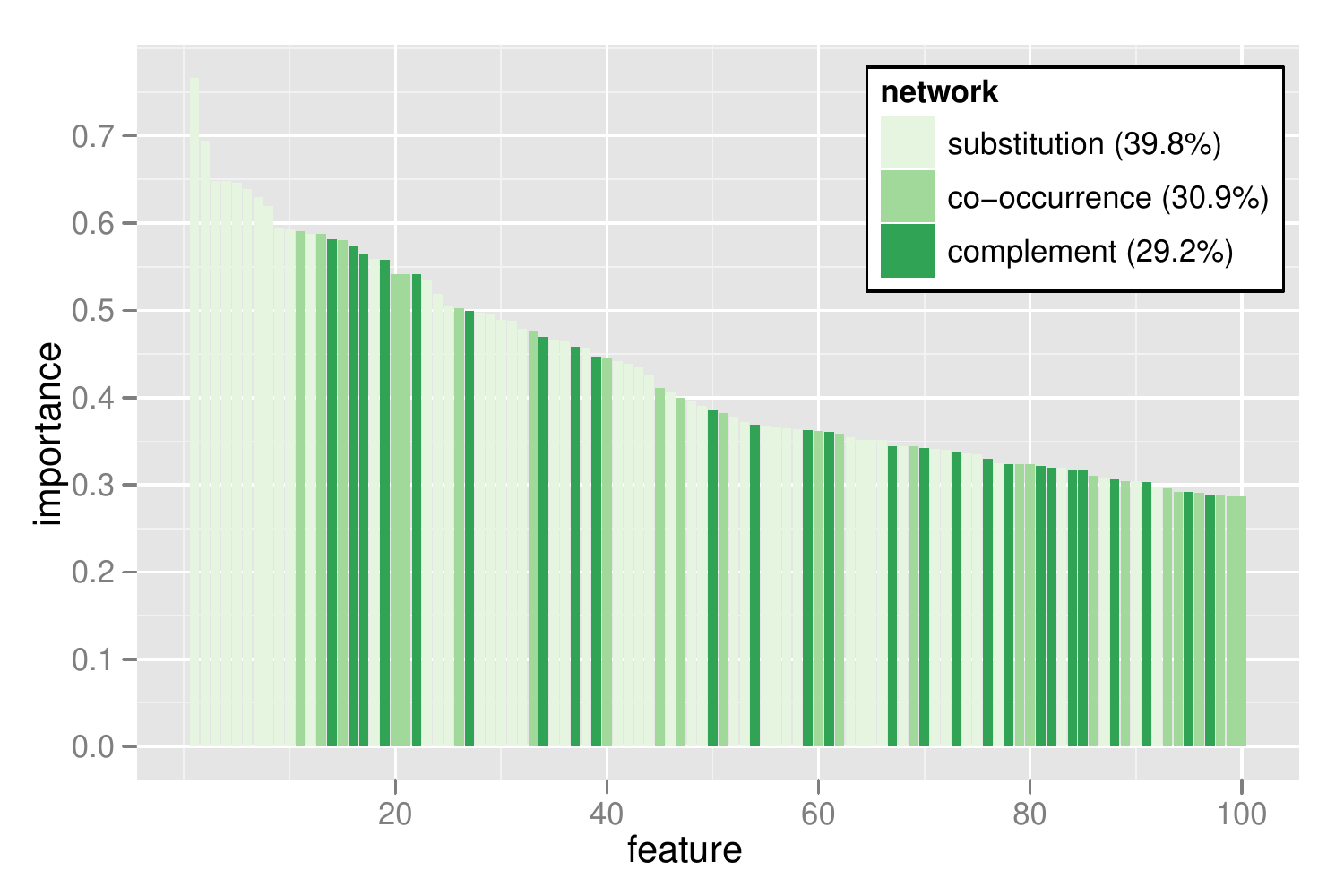}
\caption{\label{fig:relimpnet} Relative importance of features
  representing the network structure. The substitution network has the strongest contribution
($39.8\%$) to the total importance of network features, and it also has more influential features in
the top 100 list, which suggests that the substitution
network is complementary to other features.}  
\vspace{-1em}
\end{figure}

\begin{figure}[t!]\centering
\includegraphics[trim=0 30 0 0,width=1\columnwidth]{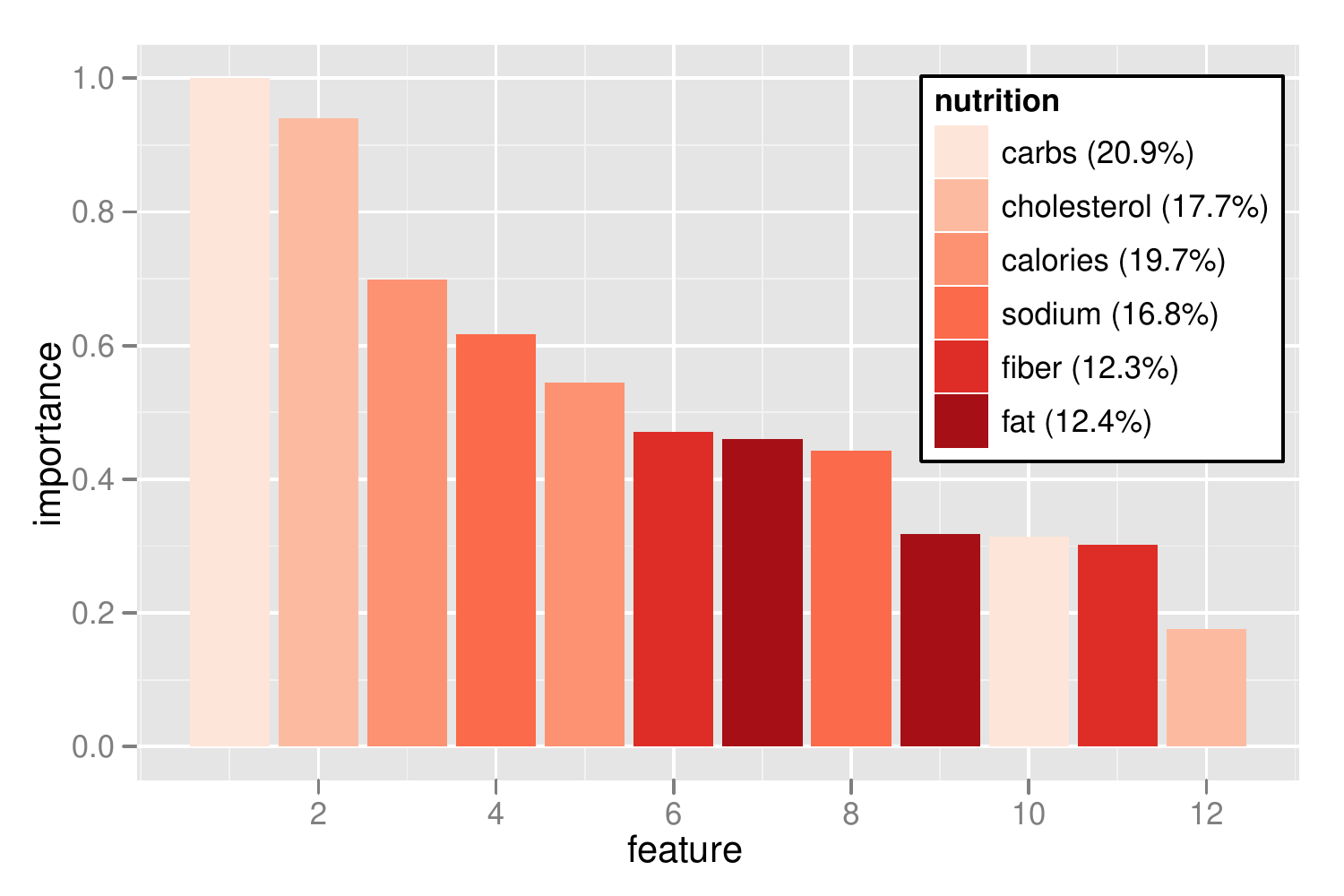}
\caption{\label{fig:relimpnut} Relative importance of features from
  nutrition information. The carbs item is the
most influential feature in predicting higher-rated recipes.}  
\vspace{-1em}
\end{figure}

The overall prediction performance is shown in
Fig.~\ref{fig:predres}. Surprisingly, even with a full list of ingredients, the
prediction accuracy is only improved from .712 (baseline)  to .746. In contrast,
the nutrition information and ingredient networks are more effective
(with accuracy .753 and .786, respectively). Both of them have much lower
dimensions (from tens to several hundreds), compared with the full
ingredients that are represented by more than 2000 dimensions (1000 ingredients per recipe in the pair). The
ingredient network features lead to impressive performance, close to
the best performance given by the combined set (.792), indicating the power of network structures
in recipe recommendation.

Figure~\ref{fig:relimp} shows the influence of different
features in the combined feature set. Up to 100 features with
the highest relative importance are shown. The importance of a feature group is
summarized by how much the total importance is contributed by all
features in the set. For example, the baseline consisting of cooking
effort and cooking methods contribute $8.9\%$ to the overall
performance. The individual items from nutrition information are very
indicative in differentiating highly-rated recipes, while most of the
prediction power comes from ingredient networks ($84\%$). 

Figure~\ref{fig:relimpnet} shows the top 100 features from the three
networks. In terms of the total importance of ingredient network
features, the substitution network has slightly stronger contribution
($39.8\%$) than the other two networks, and it also has more
influential features in the top 100 list. This suggests that
the structural information extracted from the substitution network is
not only important but also complementary to information from other aspects.

\enlargethispage{\baselineskip}
Looking into the nutrition information (Fig.~\ref{fig:relimpnut}), we found that carbohydrates are the
most influential feature in predicting higher-rated recipes. Since
carbohydrates comprise around $50\%$ or more of total calories, the
high importance of this feature interestingly suggests that a recipe's
rating can be influenced by users' concerns about nutrition and diet. Another
interesting observation is that, while individual nutrition items are
powerful predictors, a higher prediction accuracy can be reached by
using ingredient networks alone, as shown in
Fig.~\ref{fig:predres}. This implies the information about nutrition may
have been encoded in the ingredient network structure, e.g. substitutions of less healthful ingredients with ``healthier'' alternatives. 

\begin{figure}[t!]\centering
\includegraphics[trim=0 30 0 0,width=1\columnwidth]{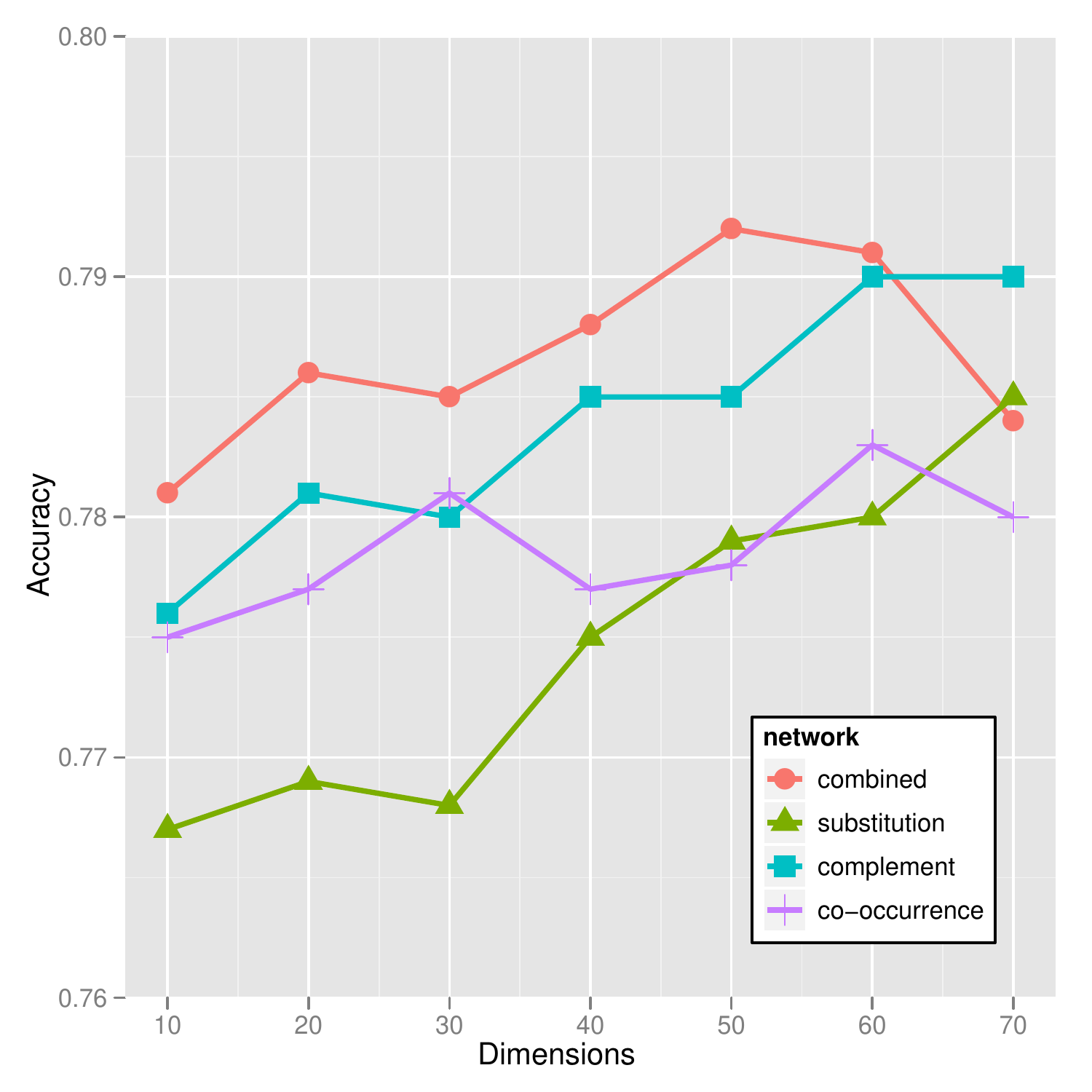}
\caption{\label{fig:dimension} Prediction performance over reduced
  dimensionality. The best performance is given by
  reduced dimension $k=50$ when combining all three networks. In
  addition, using the information about the complement network alone is more
effective in prediction than using other two networks.}  
\vspace{-1em}
\end{figure}

\begin{figure}[t!]\centering
\includegraphics[clip=true,trim=30 114 10 20,width=1\columnwidth]{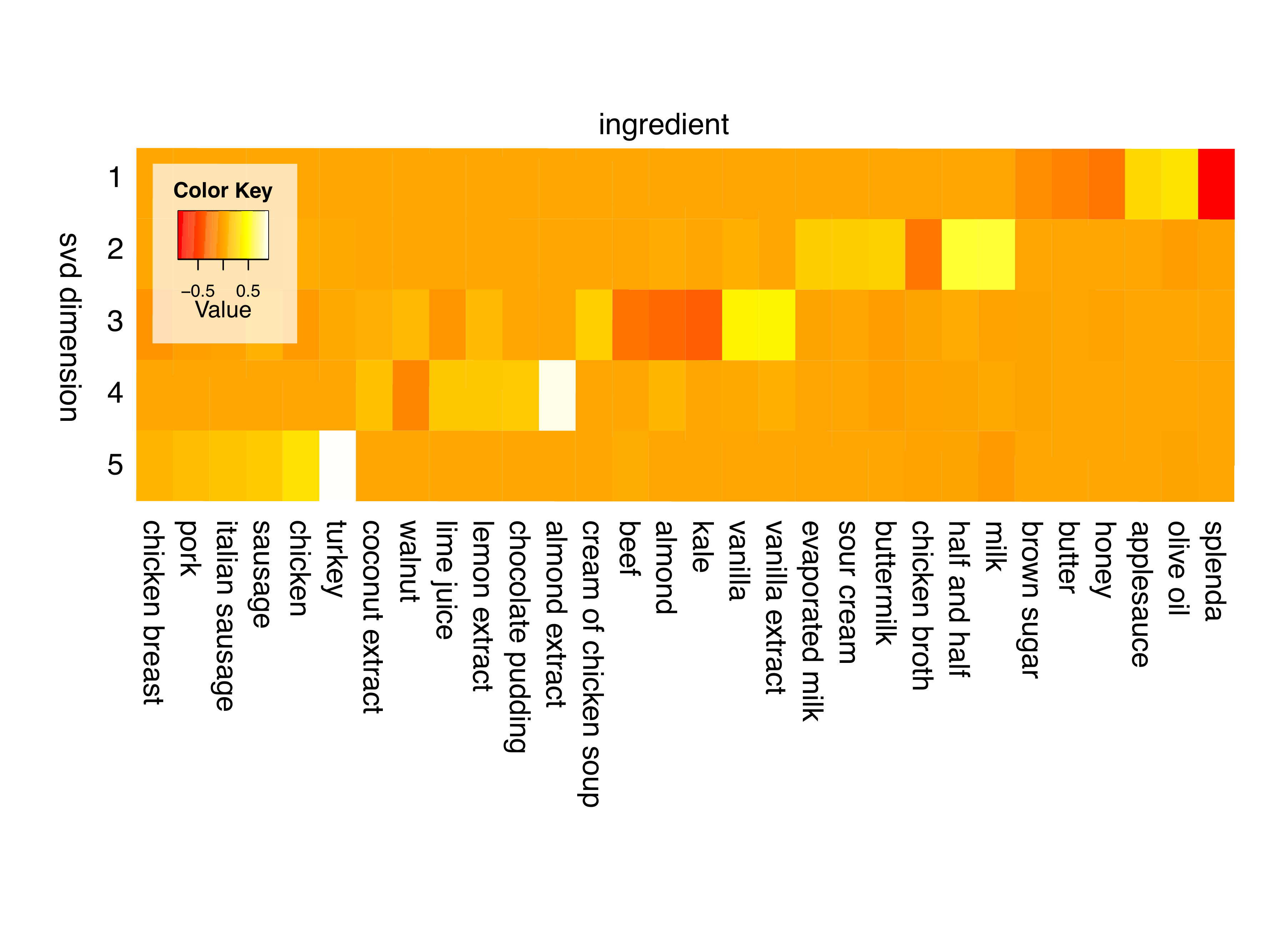}
\caption{\label{fig:svdsubs} Influential substitution communities. The matrix shows the most influential feature
  dimensions extracted from the substitution network. For each
  dimension, the six representative ingredients with the highest intensity values
 are shown, with colors indicating their
intensity. These features suggest that the communities of ingredient substitutes, such
as the sweet and oil in the first dimension, are particularly
informative in prediction. }  
\vspace{-1em}
\end{figure}

Constructing the ingredient network feature involves reducing
high-dimensional network information through SVD, as described in the previous
section. The dimensionality can be determined by cross-validation. As
shown in Fig.~\ref{fig:dimension}, features with a very large
dimension tend to overfit the training data. Hence we chose
$k=50$ for the reduced dimension of all three networks. The figure also
shows that using the information about the complement network alone is more
effective in prediction than using either the co-occurrence and substitute networks, even in the
case of low dimensions. Consistently, as shown in terms of relative
importance (Fig.~\ref{fig:relimpnet}),
the substitution network alone is not the most effective, but it provides
more complementary information in the combined feature set.

In Figure~\ref{fig:svdsubs} we show the most representative ingredients
in the decomposed matrix derived from the substitution network. We
display the top five influential dimensions, evaluated based on the
relative importance, from the SVD resultant matrix $V_k$, and in each
of these dimensions we extracted six representative ingredients based on
their intensities in the dimension (the squared entry values). These
representative ingredients suggest that the communities of ingredient
substitutes, such as the sweet and oil substitutes in the first
dimension or the milk substitutes in the second dimesion (which is
similar to the cluster shown in Fig.~\ref{fig:subclusters}), are particularly
informative in predicting recipe ratings.

To summarize our observations, we find we are
able to effectively predict users' preference for a recipe, but the
prediction is not through using a full list of ingredients. Instead,
by using the structural information extracted from the relationships
among ingredients, we can better uncover users' preference about
recipes.

\section{Conclusion}\label{sec:conclusion}
Recipes are little more than instructions for combining and processing sets of ingredients. Individual cookbooks, even the most expansive ones, contain single recipes for each dish. The web, however, permits collaborative recipe generation and modification, with tens of thousands of recipes contributed in individual websites. We have shown how this data can be used to glean insights about regional preferences and modifiability of individual ingredients, and also how it can be used to construct two kinds of networks, one of ingredient complements, the other of ingredient substitutes. These networks encode which ingredients go well together, and which can be substituted to obtain superior results, and permit one to predict, given a pair of related recipes, which one will be more highly rated by users. 

In future work, we plan to extend ingredient networks to incorporate the cooking methods as well. It would also be of interest to generate region-specific and diet-specific ratings, depending on the users' background and preferences. A whole host of user-interface features could be added for users who are interacting with recipes, whether the recipe is newly submitted, and hence unrated, or whether they are browsing a cookbook. In addition to automatically predicting a rating for the recipe, one could flag ingredients that can be omitted, ones whose quantity could be tweaked, as well as suggested additions and substitutions. 

\section{Acknowledgments}
This work was supported by MURI award FA9550-08-1-0265 from the Air Force Office of Scientific Research.
The methodology used in this paper was developed with support from funding from the Army Research Office, Multi-University Research Initiative on Measuring, Understanding, and Responding to Covert Social Networks: Passive and Active Tomography. The authors gratefully acknowledge D. Lazer for support.

\bibliographystyle{acm-sigchi}
\bibliography{recipe}

\end{document}